# Global-scale Observations and Modeling of Far-Ultraviolet Airglow During Twilight

Stanley C. Solomon[1], Laila Andersson[2], Alan G. Burns[1], Richard W. Eastes[2], Carlos Martinis[3], William E. McClintock[2], and Arthur D. Richmond[1]

[1]National Center for Atmospheric Research, High Altitude Observatory
3080 Center Green Dr., Boulder, CO 80301
Telephone: 1-303-497-2179     Email: stans@ucar.edu

[2]University of Colorado, Laboratory for Atmospheric and Space Physics, Boulder, Colorado

[3]Boston University, Center for Space Physics, Boston, Massachusetts

## Key Points

1. We analyzed measurements by the Global-scale Observations of the Limb and Disk mission of far-ultraviolet emissions near the terminator.

2. Airglow excited by photoelectrons originating in the magnetically conjugate hemisphere is observed and simulated.

3. The conjugate photoelectron source explains much of the twilight airglow, but losses occur in the plasmasphere and magnetosphere.

## Abstract

The NASA Global-scale Observations of the Limb and Disk (GOLD) ultraviolet imaging spectrograph performs observations of upper-atmosphere airglow from the sunlit disk and limb of the Earth, in order to infer quantities such as the composition and temperature of the thermosphere. To interpret the measurements, the observational and solar illumination geometry must be considered. We use forward models of upper atmosphere density and composition, photoelectron impact, airglow emissions, radiative transfer, and line-of-sight integration, to describe the expected observations, and here test those calculations against observations near the terminator, and near the limb. On the night side of the terminator, broad regions of faint airglow are seen, particularly near the winter solstice. These are caused by photoelectrons that were transported along field lines from magnetically conjugate areas in the other hemisphere, where those areas are still illuminated. We perform model calculations to demonstrate that this process is the source of the emission, and obtain good agreement with its morphology and intensity. In some regions, the observed emissions are not as intense as the model simulations. Some of the reductions in electron flux are explained by changes in magnetic field strength; in other cases, particularly at high magnetic latitude, the cause is unknown, but must occur along extended field lines as they reach into the plasma sheet.

## Plain Language Summary

The NASA Global-scale Observations of the Limb and Disk (GOLD) instrument is an ultraviolet imager and spectrograph that observes light from the upper-atmosphere of the Earth, in order to infer quantities such as the composition and temperature of the thermosphere. To interpret the measurements, the observing and solar illumination geometry must be considered. We use forward models of upper atmosphere density and composition, photoelectron impact, airglow emissions, radiative transfer, and line-of-sight integration, to describe the expected observations, and here test those calculations against observations near sunrise and sunset. At night but near twilight, broad regions of faint emissions of airglow light are seen, particularly during winter. These are caused by electrons that are created by ionization on the dayside, and are then transported along field lines from magnetically conjugate areas in the other hemisphere, where those areas are still illuminated. We perform model calculations to demonstrate that this process is the source of the emission, and obtain good agreement with the observed shape and intensity. In some regions, the observed emissions are not as intense as the model calculations. Some of the reductions in are explained by changes in magnetic field strength that affect the motions of the electrons; in other cases, particularly at high magnetic latitude, the cause is unknown, but must occur along the long paths that they travel through the magnetosphere.

## 1. Introduction

The NASA Global-scale Observations of the Limb and Disk (GOLD) instrument is an ultraviolet imaging spectrograph carried by the SES-14 communications satellite in geostationary orbit at 47.5° west longitude. It measures the Earth's far ultraviolet (FUV) airglow in the wavelength range 133 to 165 nm, using two identical spectrograph channels. The instrument images the Earth's sunlit and nightside disk, performs limb scans, and observes stellar occultations, from ~6:00 to ~24:00 universal time each day (~3:00 to ~21:00 local time at the satellite longitude), omitting only the period when the Sun is too close to the instrument field of view for detector safety. Images from each channel are constructed at a half hour cadence, twelve minutes for each hemisphere, with six minutes reserved for limb scans and occultations. Each channel has an independent scan mirror, interchangeable slits, and is capable of observing the entire disk of the Earth that is visible from geostationary orbit, ~±70° in latitude and longitude. Cross-delayline detectors provide spectral information at spatial locations along the entrance slit, and the high bandwidth provided by the communications satellite allows the detector coordinates and pulse heights of individual photon events to be downlinked; binning into pixels and all subsequent processing is done on the ground. The instrument has been described by Eastes et al. (2017) and McClintock et al. (2017).

Space-based observations of FUV airglow emissions, generally considered to be in the range 120 to 200 nm, have considerable heritage, since they are advantageous for thermosphere-ionosphere and auroral measurements because light from the lower and middle atmosphere is absorbed by molecular oxygen ($O_2$). Beginning with the iconic images from the Apollo 16 lunar mission (Carruthers et al., 1976), a long line of observations from space by missions such as OGO-4, DE-2, STP78-1, POLAR, IMAGE, TIMED, and DMSP have established the utility of FUV photometry, spectroscopy, and imaging (e.g., Hanson, 1969; Frank et al., 1982; Chakrabarti, 1984; Torr et al., 1995; Mende, 2000; Christensen et al., 2003; Comberiate & Paxton, 2010). To these, the GOLD mission adds a unique perspective from its high-altitude position in geostationary orbit, which enables it to observe a wide range of latitudes and solar times on each day. Key observational goals of GOLD are to obtain column composition ratios for atomic oxygen to molecular nitrogen, by comparing the atomic oxygen O($^5$S) doublet at 135.6 nm to molecular nitrogen Lyman-Birge-Hopfield ($N_2$ LBH) bands in the range 137 to 154 nm, and temperature from the shape of the rotational distribution of the LBH bands. In order to measure how these parameters vary with geophysical conditions, we must first take into account the changes to the emission altitude profile as a function of the solar zenith angle (SZA), and the observational geometry must also be quantified, primarily the effect of the emission zenith angle (EZA), the angle between the local vertical and the direction of the satellite. This provides a test of the ability of our models of upper atmosphere density and composition, photoelectron impact, airglow emissions, radiative transfer, and line-of-sight integration, to describe the intensity and morphology of the observations, particularly at high SZA, i.e., near the terminator, and at high EZA, i.e., near the limb.

The $N_2$ LBH bands are excited by energetic electron impact on $N_2$. Outside of the auroral zone, this is provided by photoelectrons, the super-thermal but low-energy free electrons that are produced through ionization of atmospheric gases by solar extreme-ultraviolet radiation. The O($^5$S) doublet has several possible sources, but only two are significant: energetic electron impact on O, and radiative recombination of atomic oxygen ions with ambient thermal electrons. However, broad areas of relatively faint O($^5$S) airglow are observed in the vicinity of the terminator, but well into the night, in regions outside of the auroral zone and away from the equatorial ionization anomaly (EIA), where there are no known enhancements of ionospheric density. We will demonstrate that that these emissions are excited by photoelectrons originating in the other hemisphere, that escape the atmosphere, travel along magnetic field lines to their magnetically conjugate locations, and re-enter the atmosphere, depositing their energy and causing airglow excitation, in a manner that resembles a very soft aurora.

Observations of pre-dawn airglow enhancement date back to Barbier (1959), and photoelectron transport from the magnetically conjugate hemisphere as a source of ionization was first proposed by Hanson (1963). Associated heating and airglow excitation were discussed by Cole (1965), Fontheim et al. (1968), and Duboin et al., (1968). Evidence for this process was observed in incoherent-scatter radar measurements of electron temperature by Carlson (1966, 1968) and Evans (1968). Bennett (1969), Carmen et al. (1969), Cogger & Shepherd (1969), and Smith (1969) observed twilight airglow enhancements from the ground in the 630.0 nm line, as did Christensen (1975) using the 777.4 nm multiplet. Space flight measurements of photoelectrons during conjugate point sunrise were described by Rao and Maier (1970), and Nagy & Banks (1970) performed initial photoelectron transport calculations. Conjugate excitation of ultraviolet airglow was first observed from space by the Orbiting Geophysical Observatory 4 (Meier, 1971), and the conjugate photoelectrons themselves were observed by Atmosphere Explorer C (Peterson et al., 1977), but after modeling studies by Nagy et al. (1973) showed that the conjugate photoelectron phenomenon was not a significant source of increased ionization *per se*, interest in the subject waned for some years. However, Bahsoun-Hamade et al. (1989, 1994), and Lancaster et al. (1994) measured the twilight decay of the atomic oxygen 844.6 nm triplet, and Lancaster et al. (2000) compared the twilight 844.6 nm emission to model computations that included conjugate photoelectrons. They obtained reasonable agreement from the Millstone Hill Observatory in Massachusetts, but found time-shifts between model and measurement at the Arecibo Observatory in Puerto Rico. Waldrop et al. (2007) resolved this discrepancy by using a higher-fidelity representation of the geomagnetic main field in the model calculations, employing the International Geomagnetic Reference Field (IGRF) model to obtain good agreement of the dawn and dusk time sequences with (scaled) model calculations (Richards, 2001). That work demonstrated the importance of using a realistic geomagnetic field model, and that a tilted-dipole approximation is not sufficient for accurately locating magnetically conjugate points. Finally, Richards & Peterson (2008) and Peterson et al. (2009; 2012) analyzed electron spectrometer measurements by the FAST satellite of photoelectron fluxes backscattered from the ionosphere, demonstrating that "photoelectrons are able to travel the long journey from the sunlit hemisphere to the satellite without significant degradation."

## 2. Observations of the Far-ultraviolet Airglow

### 2.1 Data Reduction and Display

Measurements shown in this paper are derived from Level 1C GOLD channel A data (version 2, revision 1), in "day" mode, which utilizes the narrowest of the three spectrograph slits, obtaining ~0.2 nm spectral resolution, full-width-half-maximum. A scan mirror moves the location of the slit across the disk from east to west, building up an image of the northern hemisphere in 12 minutes. The mirror then scans the southern hemisphere in the same fashion. This is followed by six minutes during which limb scans are performed, so full-disk images are obtained in 24 minutes, but at a 30-minute cadence. In the L1B-to-L1C processing, a small scattered light background and a variable particle radiation background are subtracted, a spectral-spatial detector flat field correction is applied, the data are converted from counts to Rayleighs (R), and binned into "superpixels" that are 0.04 nm in wavelength and 0.2°x0.2° in look direction, which projects to ~125x125 $km^2$ on the Earth's surface, at nadir. See Eastes et al. (2017; 2019), McClintock et al. (2019), and the GOLD Mission website at gold.cs.ucf.edu for further information.

We then integrate each superpixel spectrally to obtain "1356" and "LBH" brightness. For 1356, the integration interval is from 135.2 to 136.2 nm, which includes the O($^5$S) doublet at 135.6 nm and 135.9 nm, but also includes the underlying $N_2$ LBH (3,0) band at 135.4 nm. For LBH, the integration interval is from 137 to 154 nm, but excludes the atomic nitrogen triplet at 149.3 nm. The net result of the spatial-spectral binning is a sensitivity of ~1.1 R/(count/superpixel/s) for 1356 and ~1.2 R/(count/superpixel/s) for LBH. The absolute calibration is based on laboratory measurements, and should be considered preliminary at this time, as it does not yet include variations in sensitivity in the along-slit (imaging) dimension.

The methodology for making the image plots shown in this paper is to simply plot the derived brightness array as a rectangular image, and then superimpose a map projection of the latitude-longitude grid, continental boundaries, and solid Earth limb, for reference. The rigorously calculated geodetic coordinates associated with the center of each pixel, referenced to 150 km altitude, are used to calculate magnetic field coordinates and solar zenith angles, but are not otherwise employed in the image construction, i.e., there is no further geodetic binning or interpolation of the L1C data. Two consecutive scans are superimposed, one for the northern hemisphere and the following one for the southern hemisphere, to assemble each full-disk image, and a 3x3 median filter was applied to reduce the appearance of noise.

### 2.2 Excitation of the Daytime Airglow by Photoelectron Impact

Figure 1 shows two image sequences of the O 1356 and $N_2$ LBH emission for 15 October 2018, shortly after GOLD began taking data. The morning sequence, centered at 7:22, 8:22, and 9:22 UT, corresponds to local solar times at the sub-satellite point (0° latitude and 47.5° west longitude) of 4:12, 5:12, and 6:12. The evening sequence, at 20:22, 21:22, and 22:22 UT, corresponds to local solar times at the sub-satellite point of 17:12, 18:12, and 19:12. (The times given correspond to the beginning of the second scan, which is of the southern hemisphere.) Panels a-b-c and g-h-i (the first and third rows) are images of the 1356 emission, and panels d-e-f and j-k-l (the second and fourth rows) are images of the LBH emission. Both emissions are dominated by the dayglow, showing the expected gradient with SZA, increasing from the terminator toward noon, and with EZA, increasing from the sub-satellite point toward the limb, as the path length through the emitting region lengthens. The slight discontinuities noticeable near the equator are due to the 12-minute lag between the hemispheric images. The northern hemisphere aurora is also visible in both 1356 and LBH, but the southern hemisphere auroral oval is mostly behind the limb, due to the tilt and offset of the geomagnetic pole. The 1356 images exhibit additional features, including the radiative recombination nightglow from $O^+ + e$ in the equatorial ionization anomaly on either side of the magnetic equator, especially visible in the evening sequence (cf., Eastes et al., 2019), and detectable on the limb.

### 2.3 Excitation of the Twilight Airglow by Conjugate Photoelectrons

In the morning sequence, there are broad regions of faint but consistent 1356 emission in pre-dawn regions of the northern hemisphere. These features are not accompanied by significant LBH emission, but since they are not associated with any known ionospheric phenomena, we do not expect that they are caused by radiative recombination. We propose that they are caused by photoelectrons excited in the opposite hemisphere, where it is sunlit, and then transported along magnetic field lines to the emitting region. We use the term "conjugate photoelectrons" to refer to photoelectrons generated by solar EUV ionization at the magnetic field "conjugate point," the end of the magnetic field line in the opposite hemisphere. There are times and locations when and where one conjugate point is illuminated and the other is not, so conjugate photoelectrons create regions of faint nightglow. (This process occurs, of course, when both ends of the field line are illuminated, but the effects are barely discernable against the bright dayglow excited locally.) The twilight features are particularly visible in the morning sequence for 1356 shown in Figure 1. In the evening sequence, the apparent region of conjugate photoelectron excitation is much smaller, during October in the American sector, since the magnetic field lines approximately align with the terminator. However, the region of conjugate illumination becomes significant at both dawn and dusk during northern hemisphere winter solstice, as shown in Figure 2, and then switches to the southern hemisphere during northern hemisphere summer solstice, displayed in Figure 3. On these figures, the approximate location of the airglow terminator (SZA~95°) and its magnetically conjugate points in the winter hemisphere are plotted (red dashed lines), to demonstrate the morphological coherence of the proposed mechanism. All of these days, shown as examples of the twilight emissions, were at low solar activity, and were geomagnetically quiet ($K_p < 3$). Every day observed by GOLD exhibits this feature at some location and time; its morphology changes slowly with season, according to the alignment of solar illumination with the magnetic field. In the Atlantic sector, the appearance of conjugate photoelectron airglow is favored during winter mornings in the northern hemisphere and during winter evenings in the southern hemisphere, due to the negative magnetic declination.

Emissions in these regions are barely discernable in the LBH images, if at all, so it is not possible to say with certainty if conjugate photoelectrons excite the LBH bands, even using the high level of sensitivity available here. This implies a low-energy precipitation source, that primarily encounters the upper thermosphere, commensurate with our understanding of the phenomenon and the modeling described below. There are some very faint areas of apparent emission in the LBH spectral range at night, but with a pattern that is not easily discerned, and generally without corresponding 1356 enhancement. We investigated the spectral structure of some of these patches; they appear to be due to elevated detector backgrounds rather than geophysical emissions, since characteristic LBH band features were not observed.

## 3. Model Calculations

### 3.1 Model Description

Here we present a simple model of airglow emissions as seen by GOLD, extended for use with conjugate photoelectrons. The basis for our calculations is the Global Airglow (GLOW) model v. 0.983 (see Solomon et al., 1988; Solomon & Abreu, 1989; Bailey et al., 2002; Solomon, 2017). This model is based on the two-stream electron transport method of Nagy & Banks (1970); Banks & Nagy (1971). It is a single-column model, applied repeatedly on a 5°x5° latitude-longitude grid, using the solar zenith angle and magnetic field inclination appropriate for each location and time. In the calculations for this work, atmosphere and ionosphere densities, temperatures, and composition, are provided by the NRLMSISE-00 ("MSIS") model (Picone et al., 2002), the International Reference Ionosphere (IRI) (Bilitza et al., 1990), and the Nitric Oxide Empirical Model (NOEM) (Marsh et al., 2004). These empirical models provide a basic climatology that is adequate for the quiet, solar minimum conditions observed here.

After a 3-D field of the global volume emission rates is assembled, slant column brightnesses as would be observed by GOLD are calculated. A line integral is computed for each of the GOLD L1C superpixels defined in section 2.1, interpolating and integrating along a path through the array defined by the look direction of each superpixel. This simulates the image as GOLD observes it, including slant path and limb effects, without the need for geographic gridding or any approximations involved in converting vertical to slant paths. Adjustments for absorption by $O_2$ (mostly affecting the LBH emissions) and multiple-scattering by O (affecting only the O($^5$S) doublet) are made in the line integral calculation. This method can be applied repeatedly in each wavelength bin to build up complete spectra, but here are only applied at 135.6–135.9 nm, and for generic LBH bands in the 137–154 nm range. The results are image arrays for specific spectral intervals at the same resolution and registration as the data, that can be directly compared to the observations.

An additional model element employed is the conjugate photoelectron calculation. The principle is that photoelectrons traveling upward at the exobase will follow spiral trajectories along magnetic field lines. Thus, we calculate the upward flux of energetic electrons at the top level of the model at each location in the grid, apply it as the downward flux at the magnetically conjugate point, which is estimated as the point with the same magnetic longitude and the negative of the magnetic latitude, and the electron transport algorithm is run again. Then, the upward flux is calculated at the conjugate point, applied as the downward flux at the original point, and the electron transport algorithm is run a third time. This iteration is sufficient to calculate the input from the conjugate hemisphere, since the conjugate flux accounts for only a tiny fraction of the ionization in sunlit regions. This is not a new approach, since a version of it was used by Lancaster et al. (2000), but that work used a tilted-dipole approximation to the magnetic field, and it was not previously integrated into the GLOW model as a public release.

The magnetic field is based on the International Geomagnetic Reference Field (IGRF) model for epoch January 2019. Calculations are made using the quasi-dipole representation of Apex magnetic coordinates (Richmond, 1995), as implemented by Emmert et al. (2010) using a compact basis function expansion. This formulation is also used to obtain the magnetic field inclination and strength. The inclination is needed for the electron transport calculations, and the strength is used to calculate the attenuation of the flux, if any, due to changes in the magnetic field from one hemisphere to the other (see below). Figure 4a displays the magnetic latitude-longitude grid for 250 km altitude, at 15° intervals, which provides a visual guide for locating conjugate points, by following lines of constant longitude from one hemisphere to the other. The magnetic field strength is shown as a contour plot in Figure 4b; note in particular the region of weaker field in the "South Atlantic Anomaly," which actually spans regions of the South Atlantic and also much of South America. This is due to the offset and distortion of the nearly-dipole field, or, equivalently, quadrupole, octupole, and higher order terms in a spherical harmonic representation of the magnetic field.

There are some approximations involved in this method of handling conjugate fluxes, additional to those inherent in the two-stream formulation. The most important is that collisional losses in the plasmasphere are neglected. Collisions with the neutral atmosphere are insignificant above the model top (640 km), however, for long or highly inclined paths, collisions with thermal electrons could be significant. The cross section for electron-electron interaction decreases rapidly with increasing energy, so for the emissions studied here, which are primarily excited at energies $>\sim 10$ eV, electron collisional losses should be negligible. (For low-energy-threshold emissions such as the atomic oxygen ($^1$D) line at 630 nm, these could be considered.) Non-collisional changes in the flux, such as longitudinal transport due to gradient-curvature drift, etc., could also change the spatial distribution of precipitating conjugate electrons, but for a single transit through the plasmasphere, the approximation that a conjugate photoelectron will cleave to its field line is valid for the relevant spatial scale. However, pitch-angle redistribution from magnetic forces could be important, and so this effect is included in our calculations, as described below.

In earlier work, the electron flux entering the top of the model atmosphere was considered to be equal to the upward (escaping) flux from the conjugate hemisphere. Even in the absence of losses in the plasmasphere transit, there are two effects that could be significant, which we consider here: the change in cross-sectional area from one end to the other of a given "flux tube" (a small bundle of a fixed number of field lines), and pitch angle re-distribution resulting in electrons reversing direction or "mirroring," Below, we calculate the effect of each of these separately, and then combine them together.

First, we define the hemispheric flux $\Phi$ (per unit area per unit time) as the component along the field line of the differential flux $\phi$ (per unit area per unit time per steradian) integrated over a hemisphere centered on the field line. For a differential flux that is isotropically distributed over the hemisphere, $\Phi = \pi\phi$. The change in $\Phi$ due to change in the cross-sectional area of the flux tube is easily calculated from knowledge of the magnetic field strength $B$ (supplied by the main field empirical model). Since the area traversed by a flux tube is inversely proportional to field strength, and, for a constant number of electrons, the flux is inversely proportional to area, the ratio of the flux at the conjugate point $\Phi_c$ to that from the initial hemisphere $\Phi_0$ is:

$$\frac{\Phi_c}{\Phi_0} = \frac{A_0}{A_c} = \frac{B_c}{B_0} \qquad (1)$$

Where $B_0$ is the magnetic field strength at the originating point and $B_c$ is it strength at the conjugate point. Physically, this means that a fixed area at the conjugate point is drawing from a smaller or larger area at the originating point; on a regional basis, the conjugate region is either converging or diverging a larger or smaller sunlit area. Since there are regions where the change in magnetic field strength from one hemisphere to the other is significant, due to the distortion of the geomagnetic field, this can be an important effect.

The effect of pitch angle re-distribution on the electron flux re-entering the atmosphere at the conjugate point is calculated by applying conservation of the first adiabatic invariant in the collisionless regime. This dictates that the angle between an electron velocity vector and the magnetic field line, the pitch angle $\alpha$, and its initial pitch angle $\alpha_0$ as it leaves the atmosphere, are related to the magnetic field strength $B$ at any point along the field line by:

$$\frac{\sin^2 \alpha}{\sin^2 \alpha_0} = \frac{B}{B_0} \qquad (2)$$

When $B \geq B_0$, then when $\alpha$ reaches $\pi/2$ radians (90°), the electrons reverse direction or "mirror." (If $B < B_0$, $\alpha < 90°$ and none of the photoelectrons will mirror.) If $B \geq B_0$, the initial pitch angle beyond which electrons will mirror, $\alpha_m$, is:

$$\alpha_m = \sin^{-1}\left(\frac{B_0}{B}\right)^{1/2}, \quad \text{for } B \geq B_0 \qquad (3)$$

Assuming that the initial pitch angle distribution is isotropic over the upward hemisphere, the flux of electrons that will not mirror, and enter the conjugate atmosphere in the "loss cone," is:

$$\int_0^{\alpha_m} 2\pi\phi \sin\alpha \cos\alpha \, d\alpha = -\pi\phi(\cos^2 \alpha)\big|_0^{\alpha_m} = \pi\phi\left(1 - \cos^2 \alpha_m\right) \qquad (4)$$

Since the integrated along-field-line component of the unattenuated flux (the hemispheric flux) is equal to $\pi\phi$, the fraction of electrons re-entering the atmosphere at the conjugate point is:

$$\frac{\Phi_c}{\Phi_0} = 1 - \cos^2 \alpha_m = \sin^2 \alpha_m . \qquad (5)$$

Applying eqn. (3) at the conjugate point:

$$\frac{\Phi_c}{\Phi_0} = \frac{B_0}{B_c}, \quad \text{for } B_c \geq B_0, \qquad (6)$$

$$\frac{\Phi_c}{\Phi_0} = 1, \quad \text{for } B_c < B_0$$

Therefore, when the two effects described by eqns. (1) and (6) are combined, for $B_c \geq B_0$ (a converging flux tube), eqns. (1) and (6) cancel, and there is no change in flux. However, for $B_c < B_0$ (a diverging flux tube) the net effect is simply equation 1:

$$\frac{\Phi_c}{\Phi_0} = 1, \quad \text{for } B_c \geq B_0 \qquad (7)$$

$$\frac{\Phi_c}{\Phi_0} = \frac{B_c}{B_0}, \quad \text{for } B_c < B_0$$

Note that these simple expressions only hold for hemispherically isotropic fluxes. In the two-stream electron transport formulation, the fluxes are assumed to be isotropic over each hemisphere, but are then approximated by a characteristic pitch angle of $\pi/3$ radians (60°). In reality, the pitch angle distribution of the exit flux will be slightly enhanced in the along-field-line direction, because electrons travelling a shorter path length are less likely to collide before escaping. We neglect this effect in the calculations shown here, and only multiply the exiting fluxes by eqn. (7) before applying them as entering fluxes at the upper boundary of the atmosphere at the conjugate points.

GLOW model calculations are performed at every altitude level at every grid point on the globe. Then, line-of-sight observing geometry is applied to compute the emergent radiance at every angle and every wavelength that GOLD observes, by integrating through the emitting region of the atmosphere. This is done on the same 0.2° x 0.2° x 0.04 nm observational grid as the GOLD L1C data. The effect of absorption by $O_2$ is included, and a simple radiative transfer approximation is applied to estimate the effect of resonant scattering by O on the 135.6 nm doublet. The LBH radiances are calculated in the 137–154 nm range, and the (3,0) band at 135.4 nm is included in the simulations of "1356," as it is in the observations. In Figure 5, we present an example of model calculations for 19 December 2018, at 7:22 UT. Note that auroral excitation is omitted from these runs, but nominal recombination emission is simulated, based on a nominal IRI ionosphere, which is visible near the equator and in the nighttime northern hemisphere. The conjugate photoelectrons primarily excite atomic oxygen, since they deposit their energy where it is the dominant species, in the altitude range 200–300 km, but, in these simulations, there is some $N_2$ LBH excitation as well. Model runs are shown with (5*a,d*) and without (5*b,e*) the magnetic field effects included, and the difference between them (5*c,f*). This demonstrates that the magnitude of these effects is modest in most regions, but becomes prominent over the South Atlantic, where the magnetic field is weaker than in the conjugate regions, reaching a maximum reduction of ~40%.

**3.2 Comparison of Model to Measurements**

The model simulations shown include the magnetic field effects as described above. Figure 6 displays the measurement, model simulation, and normalized differences (model-measurement)/measurement) for the 19 December 07:22 case from Figure 2, where that time-tag corresponds to the end of the northern hemisphere scan and the start of the southern hemisphere scan, i.e., the midpoint of the two scans. However, for these simulations, the model was run twice, for the northern hemisphere at 07:13, and for the southern hemisphere at 07:25, to account for the 12-minute lag between scans and to align the model time with the approximate location of the terminator, where the emission rate is changing rapidly. The 3x3 median filter was applied to the observation, as above, and regions where the observed brightness was less than 2 R were excluded from the difference plot (set to zero difference), for clarity, since measurements below that threshold are not considered statistically significant.

The overall morphology of the simulated 1356 emission is very similar to the observations, although the model appears to be a little low in the sunlit region, and yields a small amount of southern hemisphere recombination nightglow (based in this case on IRI for the ambient ionosphere) that is not detectable in the observations, other than on the limb. Also, we note again that there is no auroral excitation included in these model runs. The shape of the conjugate photoelectron excitation region is captured with good fidelity, but there is a significant discrepancy in the northwest area of the sub-auroral region near 60° magnetic latitude, which is an "*L*-shell" of ~4, where the model overestimates observed airglow.

Model LBH emissions are in good agreement with observations in the sunlit region. The model predicts very weak conjugate excitation of LBH as well, although they encounter a predominantly atomic oxygen atmosphere. LBH emissions are essentially absent in observations of the conjugate photoelectron region. There are some patchy spots near mid-latitudes at night that are not seen in the model; since these have no geographic or geomagnetic consistency, and are just a few Rayleighs at most, we take them to be detector noise rather than geophysical emissions. There is also a small "fringe" at the night edge of the terminator in the difference plots, where the emission is also very low, and any slight error could account for the discrepancy. Possible candidates are the simulation virtual time, solar illumination calculation, or layer height. This does not appear to be a problem in the 1356 differences, where the terminators are not as sharp.

In order to be more quantitative with these model-data comparisons, we show line plots of zonal variations through the observed and simulated images, at ±30° latitude. Results for the December solstice case are displayed in Figure 7. These confirm that the model 1356 emission is slightly lower than observed in the sunlit region, particularly in the southern (summer) hemisphere, with fairly consistent variation with solar zenith angle. This is within the current calibration uncertainty, but could also be due to seasonal/hemispherical variation in thermospheric composition that is only partially captured by the MSIS model atmosphere that underlies these airglow simulations. The terminator decay and conjugate excitation region are in good agreement. The LBH curves show excellent agreement except for right at the terminator, as seen in the images, but the discrepancy is only significant below ~20 R, or about 2 R per band, which may be falling below the observational threshold.

## 4. Discussion

The approximation that the photoelectron flux entering the "top" of the thermosphere is equal to that exiting the magnetically conjugate hemisphere, after adjustment for magnetic field effects, appears to be valid for a wide range of magnetic latitude. However, it breaks down at high magnetic latitude, where the paths are very long, and the photoelectrons travel almost through the outer radiation belt. In these regions, there must be unaccounted attenuation occurring along the higher-latitude field lines, since the model overestimates what is observed. This could be due to electron scattering losses, additional pitch-angle redistribution, field-aligned currents, meridional transport, or some combination. We also note that, on the night side, as the auroral region is approached, the dipole-like configuration of the magnetic field starts to distort into the plasma sheet and magnetotail. Therefore, we do not consider the conjugate photoelectron model to be valid > ~60° magnetic latitude or > $L$~4. The model captures the overall global morphology nonetheless, and the well-defined boundaries corresponding to the conjugate terminators clearly confirm that the source region is defined by the illuminated conjugate loci.

In the portion of the images that are directly illuminated by the Sun, the GLOW model does an excellent job of describing the combined effects of solar zenith angle and observation angle as they change throughout the field of view, in addition to any changes in atmospheric composition. There are some offsets in absolute magnitude, which could be due to several factors. In particular, since the instrument sensitivity calibration is preliminary at this time, it would be premature to place too much emphasis on absolute intensity offsets. The model cross sections and solar inputs are the same as used in Solomon (2017), which obtained reasonable agreement with limb brightnesses observed by TIMED/GUVI (Meier et al., 2017), and hence indirectly bridge the two instruments. It should be pointed out, however, that the model parameters used by Meier et al. are not identical to ours (see Solomon (2017) for further discussion). The effective cross section for O($^5$S3s) excitation is complicated by cascade contributions through the $^5$P3p state, which decays to $^5$S3s via the 777.4 nm transition. This emission has been measured from the ground, (e.g., Christensen, 1975) and, together with the similar 844.6 nm feature, forms the most direct basis for comparison with space-based observations, since the brighter 630 nm line has a lower energy excitation cross section, and a very different chemical source in the recombination nightglow.

LBH is virtually undetectable in the conjugate photoelectron regions at night, although the model does predict a very weak presence. This is because the ~10~30 eV photoelectrons are so low in energy (compared to, for instance, primary auroral electrons) that they do not penetrate below ~200 km, and hence are absorbed in a neutral atmosphere dominated by atomic oxygen. The model finds that the peak absorption/excitation region of these electrons is at ~250 km (during low solar activity) for all emissions, with rapid reduction with decreasing altitude below. There are some faint and noisy LBH features that become evident in the logarithmically-scaled images, but these have no apparent geographic or geomagnetic consistency, and no corresponding oxygen emission. Measurement of the LBH emission relies on integration across a much broader spectral range, and is thus more subject to contamination from radiation or scattered light backgrounds. Nevertheless, it is possible that high-energy particle fluxes (that would not cause much oxygen excitation) could cause these irregular features, so we do not exclude the possibility that they could be geophysical rather than instrumental, even though they are extremely weak, only a few Rayleighs at most.

The oxygen excitation is weak as well, although easily detectable on every night by the extraordinarily sensitive GOLD imager. As originally assessed by Nagy et al (1973), the energy deposited by this process is not a significant contribution to thermosphere-ionosphere energetics, although it could affect electron heating, which would be preferentially enhanced by a low-energy electron flux. We report on the phenomenon regardless, since one purpose of this study is to establish a methodology for distinguishing between electron-impact and radiative recombination sources of $O(^5S)$ excitation and 135.6 nm emission. Recombination nightglow dominates the nightside portions of the evening images (see Eastes et al., 2019), but is not the focus here, which is why a simple empirical model for the ionosphere can be used in the model simulations.

## 5. Conclusions

This is only the second report of space-based observations of twilight airglow excited by photoelectrons emanating from the magnetically conjugate areas in the opposite hemisphere, but it is the first to produce synoptic images of the phenomenon. The extensive ground-based evidence (at visible-light wavelengths) can obtain temporal dependence, but these single-point observations yield a limited window into the global morphology. The ultraviolet emissions are faint, but are similar in intensity to recombination airglow produced by an entirely different mechanism, the radiative recombination of $O^+$ with electrons, which is important for imaging the night ionosphere. So, it is necessary to distinguish between these sources on the basis of location, morphology, and theory. Near the terminators, the complex distribution of direct and indirect excitation is adequately described by model calculations, except that at the higher magnetic latitudes ($L>\sim4$), where the assumption that what goes out must come back in does not fully hold along very extended magnetic field lines. At magnetic mid-latitudes, calculations of flux tube size and pitch angle redistribution in the presence of changing magnetic field strength appears to account for processes that diminish the conjugate flux. Mirrored electrons are probably not "trapped," however, as they will likely be thermalized by collisions in the hemisphere where they originated.

These observations and associated modeling were all performed under solar minimum conditions, and geomagnetically quiet days were chosen for display here, although the conjugate nightglow is seen on all nights. We do not expect that higher solar activity will cause the morphology to be very different, since the direct and conjugate excitation should be proportional, but it could be brighter, unless increased plasmasphere attenuation offsets the increased source term. That may help to resolve whether some of the apparent features in the LBH observations are actually geophysical, or just instrument noise or background artifacts. The 1356 doublet, being a much sharper spectral feature, is not as subject to detector background issues, and is a sensitive indicator of thermosphere, ionosphere, and even magnetosphere processes in these global-scale observations.

*Acknowledgements*. GOLD data are available at http://gold.cs.ucf.edu and at the NASA Space Physics Data Facility. The GLOW model is an open-source code available under an Academic Research License, from NCAR, at https://www2.hao.ucar.edu/modeling/glow, or through GitHub. This research was supported by NASA contract 80GSFC18C0061 to the University of Colorado, and by the National Center for Atmospheric Research, which is a major facility sponsored by the National Science Foundation under cooperative agreement 1852977.

# Figures

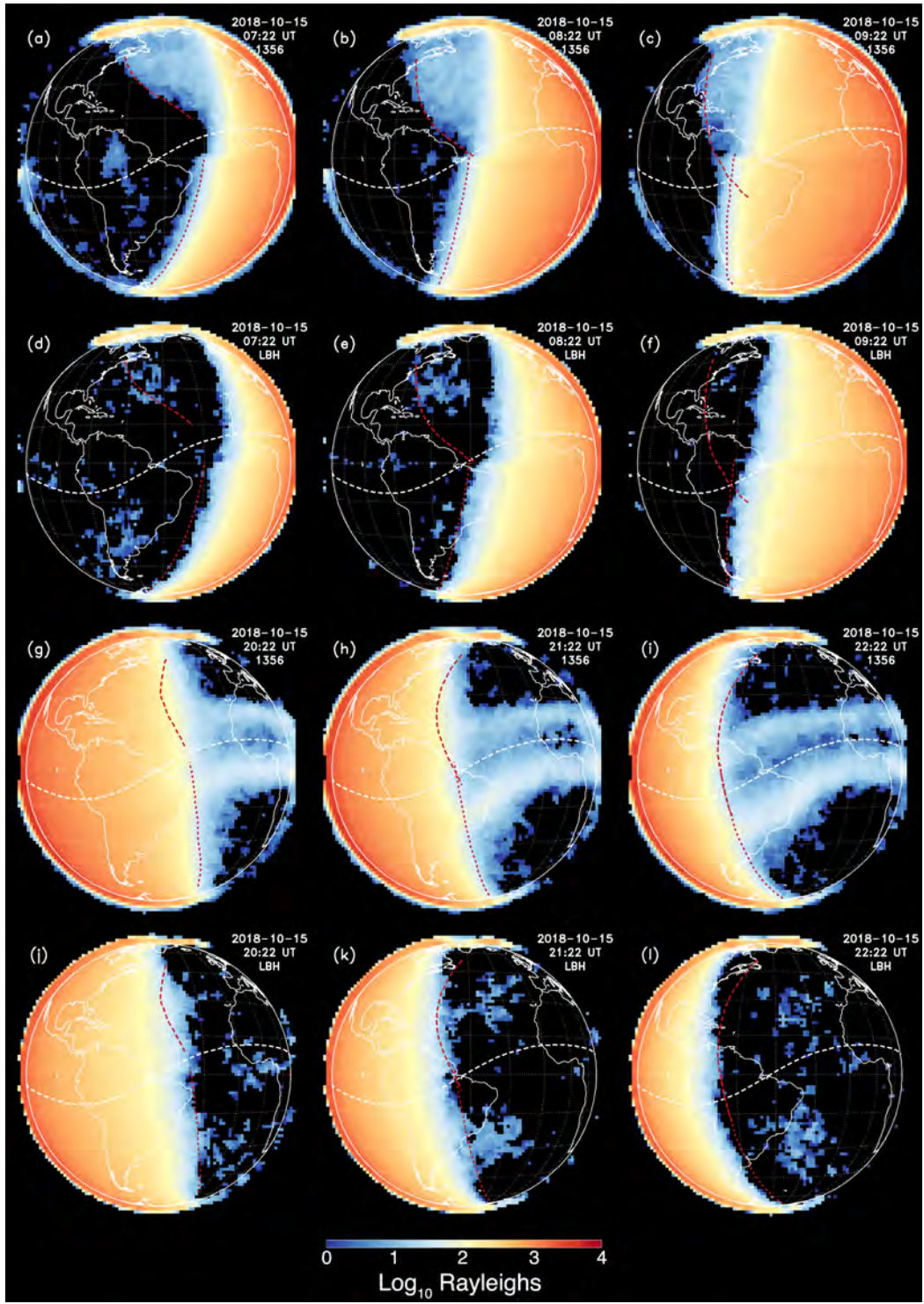

**Figure 1.** Ultraviolet emissions observed by GOLD on 15 October 2018, in the OI ($^5$S) doublet at 135.6 nm (a-c; g-i) and N$_2$ LBH bands at 137–154 nm (d-f; j-l), for morning (a-f) and evening (g-l) sequences. On the nightside, the morning 135.6 sequence shows evidence of conjugate photoelectron excitation, while the evening sequence is dominated by recombination emission in the EIA. These features are mostly absent in the LBH bands. The magnetic equator is shown by the white dashed line; the southern hemisphere terminator, and its conjugate trace, are shown by dotted and dashed red lines, respectively.

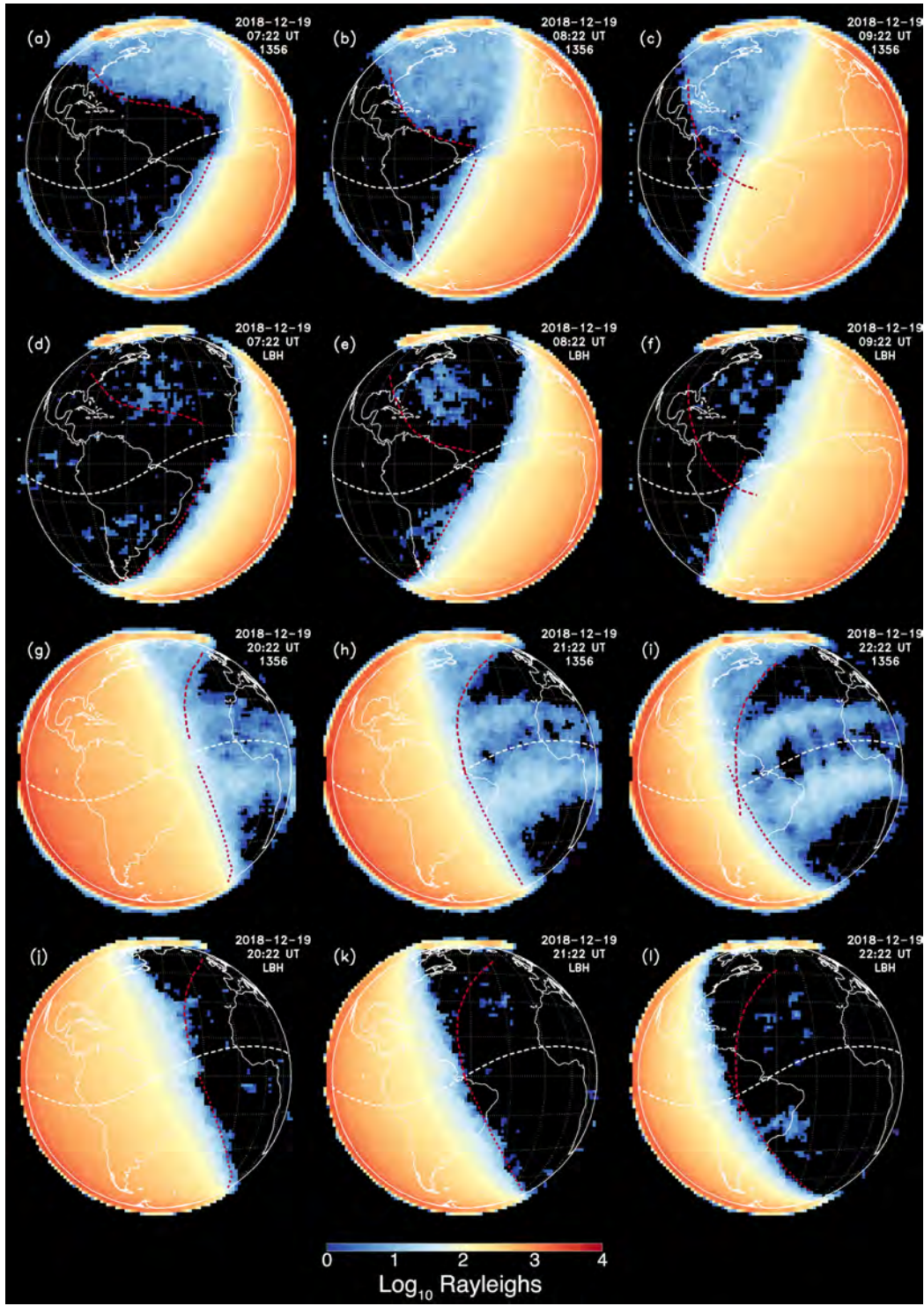

**Figure 2.** Ultraviolet emissions observed by GOLD on 19 December 2018, in the OI ($^5$S) doublet at 135.6 nm (a-c; g-i) and N$_2$ LBH bands at 137–154 nm (d-f; j-l), for morning (a-f) and evening (g-l) sequences. The nightside conjugate photoelectron excitation area during Northern hemisphere morning is larger during winter solstice, and is also visible in the evening sequence. Irregular depletion regions can be detected in the EIA, even using this low-sensitivity slit. The magnetic equator is shown by the white dashed line; the southern hemisphere terminator, and its conjugate trace, are shown by dotted and dashed red lines, respectively.

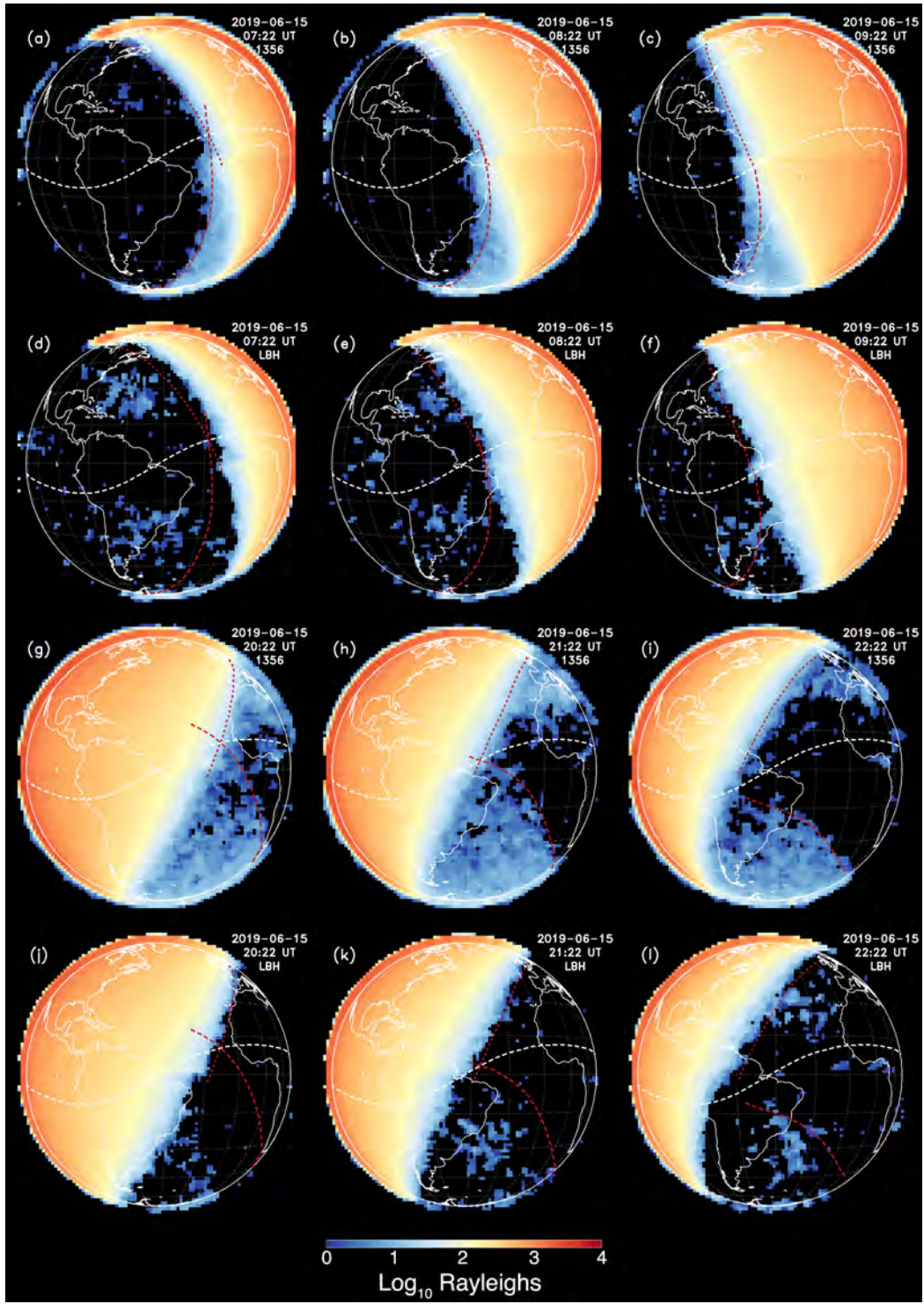

**Figure 3.** Ultraviolet emissions observed by GOLD on 15 June 2019, in the OI ($^5$S) doublet at 135.6 nm (a-c; g-i) and $N_2$ LBH bands at 137–154 nm (d-f; j-l), for morning (a-f) and evening (g-l) sequences. The nightside conjugate photoelectron excitation area during Southern hemisphere morning is smaller than at the opposite solstice, but larger in the evening sequence. The magnetic equator is shown by the white dashed line; the northern hemisphere terminator, and its conjugate trace, are shown by dotted and dashed red lines, respectively.

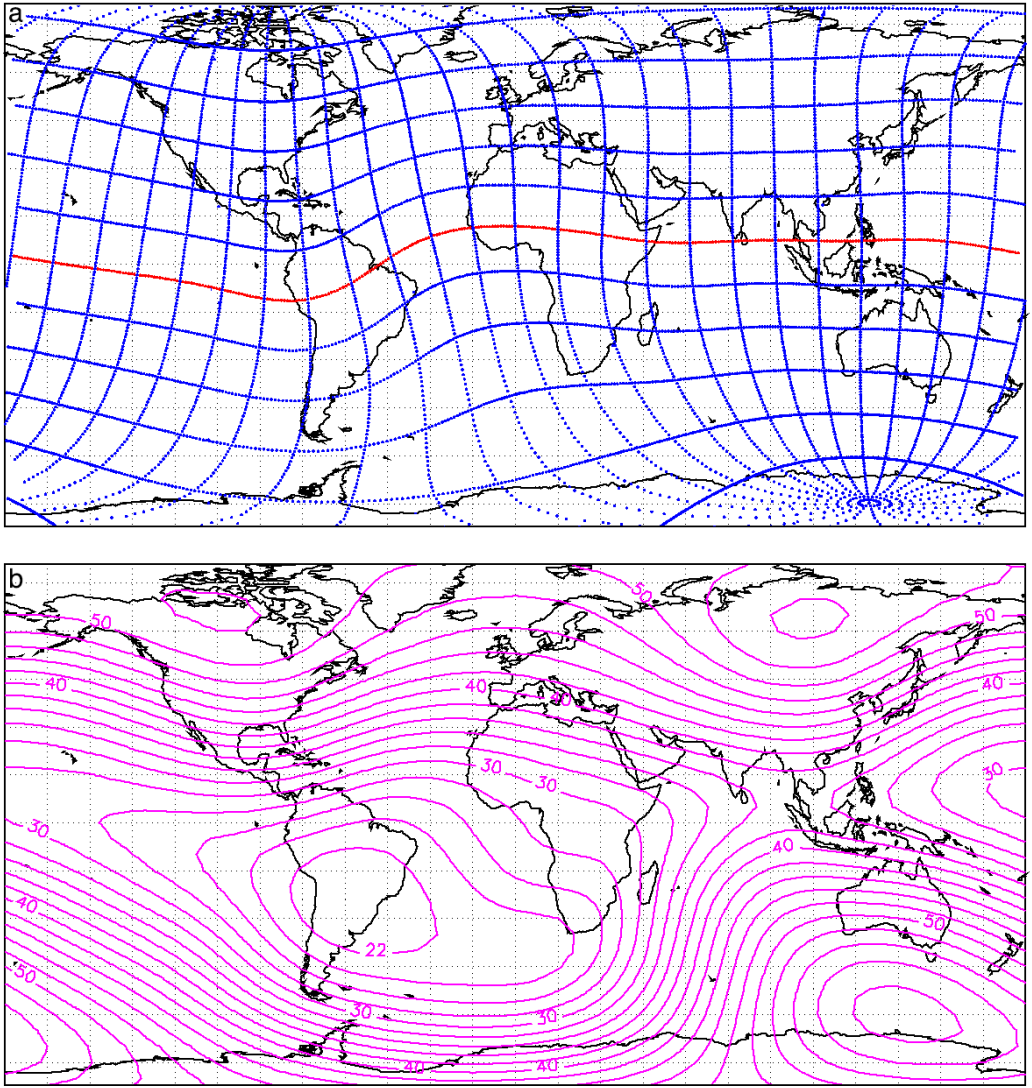

**Figure 4.** (a) Lines of equal quasi-dipole magnetic latitude and longitude (at 250 km altitude) on the geodetic grid, based on the IGRF magnetic field model for 1 January 2019. The red line denotes the magnetic equator. (b) Contours of the magnetic field strength in microTesla (µT). Both plots extend to latitude ±80° and longitude ±180°.

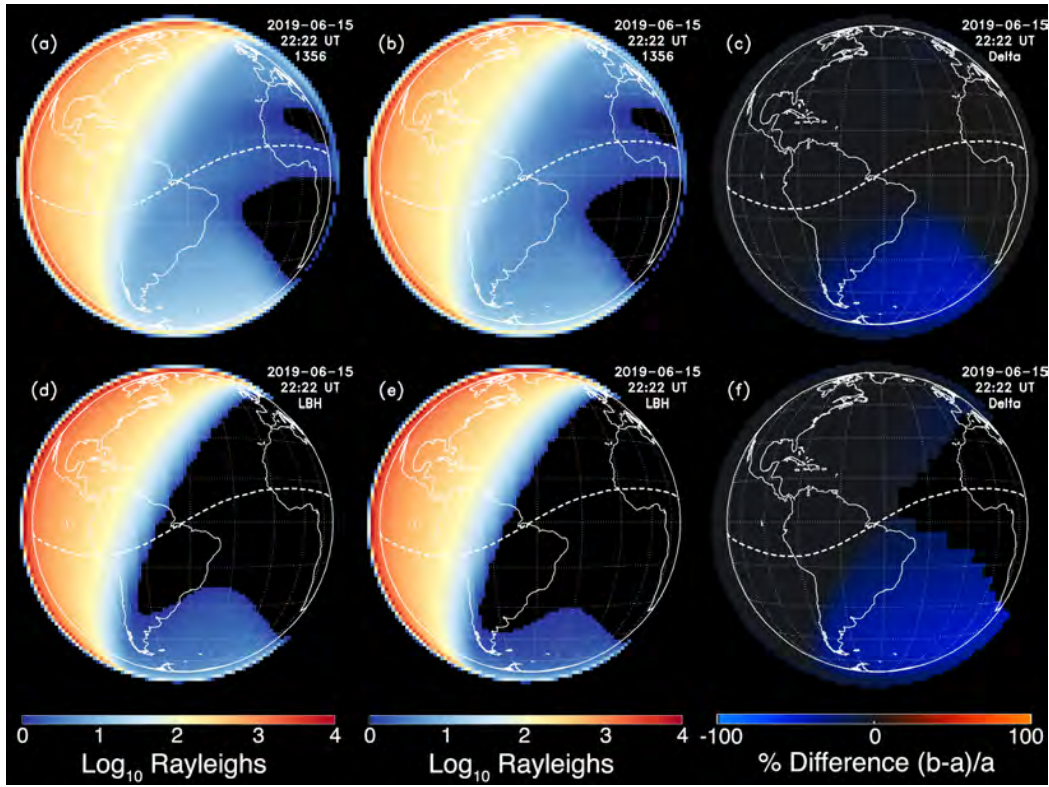

**Figure 5.** Model simulations of emission brightness as observed from geostationary orbit, for 15 June of 2019. Top row (a-c): O I 135.6 nm doublet. Bottom row (d-f): $N_2$ LBH bands 137–154 nm. Left panels (a and d): model slant column brightness without the effects of flux tube dilation on conjugate photoelectrons. Center panels (b and e): model with flux tube dilation included. Right panels (c and f): Percentage difference between the two simulations, (with - without) / without.

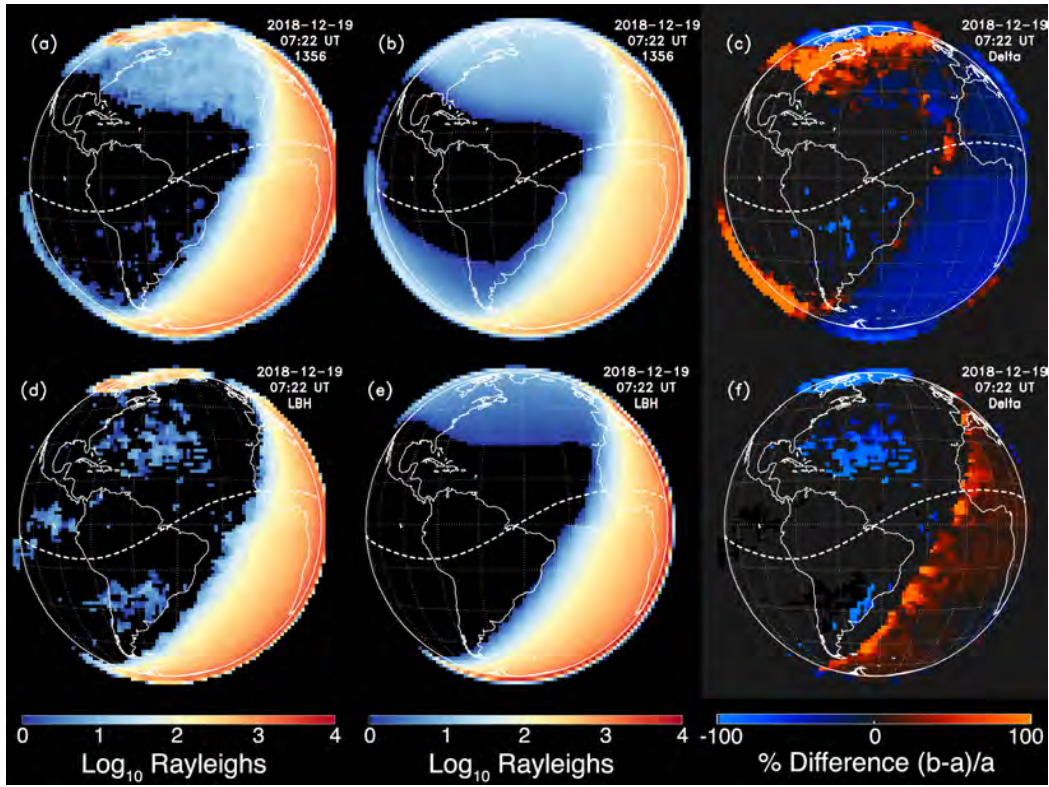

Figure 6. Measurement, model simulation, and normalized differences for the 19 December case at 07:22 UT. Top row (a-c): O I 135.6 nm doublet. Bottom row (d-f): $N_2$ LBH bands 137–154 nm. Left panels (a and d): observed slant column brightness. Center panels (b and e): Model with magnetic effects included. Right panels (c and f): Percentage difference (model - measurement) / measurement). For these simulations, the model was run twice, for the northern hemisphere at 07:13, and for the southern hemisphere at 07:25, to account for the 12-minute lag between scans, and to align the model time with the approximate location of the terminator. Areas where the observation was less the 2 R were excluded from the difference display.

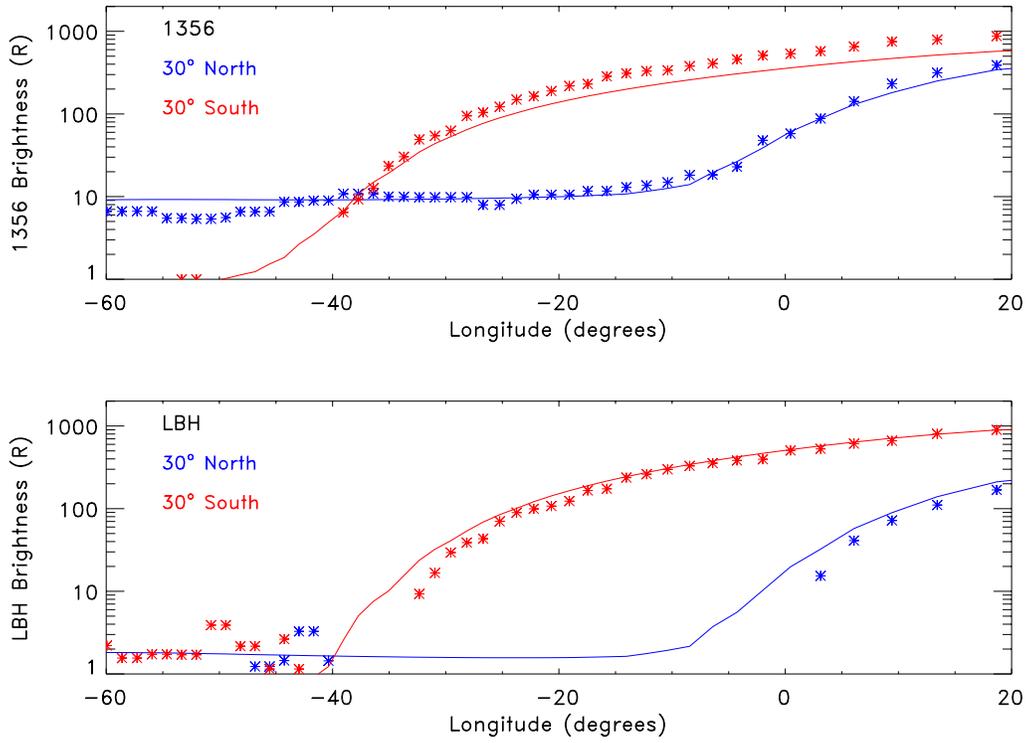

Figure 7. Comparison of observations to model for the 19 December case at 07:22 UT shown in Figure 6, in two rows of image pixels located at 30° north latitude and 30° south latitude. Top: O I 135.6 nm doublet. Bottom: $N_2$ LBH bands 137–154 nm. Lines: model simulation. Symbols: GOLD measurement. Blue: 30° north latitude. Red: 30° south latitude.